\overfullrule=0pt
\input harvmac

\def\a{\alpha}

\def\b{\beta}
\def\g{\gamma}
\def\l{\lambda}

\def\d{\delta}
\def\t{\theta}

\def\s{\sigma}
\def\o{\omega}

\def\O{\Omega}
\def\Ob{\overline\Omega}
\def\S{\Sigma}

\def\L{\Lambda}

\def\N{\nabla}
\def\Nb{\overline\nabla}
\def\Pb{\overline\Pi}

\def\p{\partial}
\def\pb{\overline\partial}

\def\Jb{\overline J}
\def\half{{1\over 2}}

\def\Ab{\overline A}
\def\e{\epsilon}
\def\td{\tilde\delta}

\Title{ \vbox{\baselineskip12pt \hbox{}}} {\vbox{\centerline{A Note
on the Classical BRST Symmetry of the}
\smallskip
\centerline{Pure Spinor String in a Curved Background}}}

\smallskip
\centerline{Osvaldo Chand\'{\i}a\foot{e-mail: ochandia@unab.cl}}
\smallskip
\centerline{\it Departamento de Ciencias F\'{\i}sicas, Universidad
Andr\'es Bello} \centerline{\it Rep\'ublica 252, Santiago, Chile}

\bigskip

\noindent

The classical pure spinor version of the heterotic superstring in
a supergravity and super Yang-Mills background is considered. We
obtain the BRST transformations of the world-sheet fields. They
are consistent with the constraints obtained from the nilpotence
of the BSRT charge and the holomorphicity of the BRST current.


\Date{April 2006}

\newsec{Introduction}

A covariant formalism of the superstring was formulated six years
ago by Berkovits \ref\berk{N.~Berkovits, ``Super-Poincare
Covariant Quantization of the Superstring,'' JHEP 0004 (2000) 018
[arXiv:hep-th/0001035]. }. Since then, this formalism has passed
many tests which include the  calculation of tree-level
\ref\tree{N.~Berkovits and B.~C.~Vallilo, ``Consistency of
Super-Poincare Covariant Superstring Tree Amplitudes,'' JHEP 0007
(2000) 015 [arXiv:hep-th/0004171].} and higher loops \ref\loops{
N.~Berkovits, ``Multiloop Amplitudes and Vanishing Theorems Using
the Pure Spinor Formalism for the Superstring,'' JHEP 0409 (2004)
047 [arXiv:hep-th/0406055]. } scattering amplitudes. It was also
proven that the formalism describes correctly the superstring
degrees of freedom, in fact the superstring spectrum was
determined in the light-cone gauge \ref\lc{N.~Berkovits,
``Cohomology in the Pure Spinor Formalism for the Superstring,''
JHEP 0009 (2000) 046 [arXiv:hep-th/0006003]\semi N.~Berkovits and
O.~Chand\'{\i}a, ``Lorentz Invariance of the Pure Spinor BRST
Cohomology for the Superstring,'' Phys.\ Lett.\ B514 (2001) 394
[arXiv:hep-th/0105149].} and in \ref\mass{N.~Berkovits and
O.~Chand\'{\i}a, ``Massive Superstring Vertex Operator in D = 10
Superspace,'' JHEP 0208 (2002) 040 [arXiv:hep-th/0204121]. } it
was constructed the first massive state in terms of a manifestly
ten dimensional supercovariant language. Recently, Berkovits
realized that his formalism admits a more geometrical origin by
discovering a topological formulation \ref\top{ N.~Berkovits,
``Pure Spinor Formalism as an N = 2 Topological String,'' JHEP
0510 (2005) 089 [arXiv:hep-th/0509120]. }.

The formalism can be adapted to describe strings in curved
backgrounds including those with Ramond-Ramond fluxes like anti de
Sitter spaces \ref\ads{N.~Berkovits and O.~Chand\'{\i}a,
``Superstring Vertex Operators in an AdS$_5\times$S$^5$
Background,'' Nucl.\ Phys.\ B596 (2001) 185
[arXiv:hep-th/0009168].}. There, quantum conformal invariance
\ref\vall{B.~C.~Vallilo, ``One Loop Conformal Invariance of the
Superstring in an AdS$_5\times$S$^5$ Background,'' JHEP 0212
(2002) 042 [arXiv:hep-th/0210064].} and quantum BRST invariance
\ref\bb{ N.~Berkovits, ``Quantum Consistency of the Superstring in
AdS$_5\times$ S$^5$ Background,'' JHEP 0503 (2005) 041
[arXiv:hep-th/0411170].} have been verified.

Berkovits and Howe constructed the sigma model action suitable to
describe ten dimensional supergravity backgrounds
\ref\bh{N.~Berkovits and P.~S.~Howe, ``Ten-dimensional Supergravity
Constraints from the Pure Spinor Formalism  for the Superstring,''
Nucl.\ Phys.\ B635 (2002) 75 [arXiv:hep-th/0112160].} (see also
\ref\OdaZM{I.~Oda and M.~Tonin, ``On the Berkovits Covariant
Quantization of GS Superstring,'' Phys.\ Lett.\ B520 (2001) 398
[arXiv:hep-th/0109051].}). The sigma model action is the most
general classically conformal invariant compatible with the
isometries of the background. The classical BRST invariance of the
model implies that the background fields are constrained to satisfy
the ten dimensional supergravity equations of motion. In \ref\cv{
O.~Chand\'{\i}a and B.~C.~Vallilo, ``Conformal Invariance of the
Pure Spinor Superstring in a Curved Background,'' JHEP 0404 (2004)
041 [arXiv:hep-th/0401226].} it was shown that the conformal
invariance is preserved in the quantum regime at the one-loop level
if the background is constrained by classical BRST invariance. The
next logical step is to preserve quantum BRST invariance to obtain
$\a'$-corrections in the supergravity equations of motion. In this
calculation it will be useful to determine how the world-sheet
fields transform under classical BRST transformations. The purpose
of this paper is to determine such transformations.

In the next section we review the sigma model action for the
heterotic string in the pure spinor formalism. In section 3 we
derive the classical BRST transformations of the world-sheet
fields\foot{These transformations are also obtained in
\ref\gutt{Sebastian Guttenberg, private communication.}.}. In the
final section we find consistency with the constraints of \bh\
derived from the nilpotence of the BRST charge and the
holomorphicity of the BRST current.

\newsec{The Pure Spinor Approach to the Heterotic Superstring}

Let us remind the sigma model action for the heterotic string in a
background supporting gauge and gravitational fields, it is given
by

\eqn\action{\eqalign{S={1\over{2\pi\a'}}&\int
d^2z~[\half\Pi^a\Pb^b\eta_{ab}+ \half\Pi^A\Pb^B B_{BA}+\Pi^A\Jb^I
A_{IA}\cr +d_\a(\Pb^\a+\Jb^I
W^\a_I)&+\l^\a\o_\b(\Pb^A\O_{A\a}{}^\b+\Jb^I
U_{I\a}{}^\b)]+S_{\Jb}+S_{\l,\o}+S_{FT},\cr}} where $\Pi^A=\p Z^M
E_M{}^A, \Pb^A=\pb Z^M E_M{}^A$ with $E_M{}^A$ being the
supervielbein, $Z^M=(x^m,\t^\mu); m=0,\dots, 9, \mu=1,\dots, 16$
the superspace coordinates, $d_\a$ the world-sheet generator of
superspace translations. $S_{\Jb}$ is the action for the gauge
group variables, $S_{\l,\o}$ is the action for the pure spinor
variables $(\l,\o)$\foot{Since the pure spinor $\l$ is constrained
to satisfy $\l\g^a\l=0$, its canonical conjugate $\o$ is defined
up to $\d\o_\a=(\g^a\l)_\a\L_a$ for a parameter $\L$. Then,
$\l^\a\o_\b$ can only be expressed in terms of the ghost number
current $J=\l^\a\o_\a$ and the generator for pure spinors Lorentz
rotations $N^{ab}=\half(\l\g^{ab}\o)$.} and $S_{FT}$ is the
Fradkin-Tseytlin which is proportional to $\int d^2z r \Phi$,
where $r$ is the world-sheet curvature and the $\t$-independent
term of the superfield $\Phi$ is the dilaton. This term is not
necessary when we study the classical dynamics of the system.
However, it helps to restore the quantum conformal invariance as
it was shown in \cv. The other background fields in \action\ are
the 2-form superfield $B$, the gauge field $A_I$, the superfields
$W^\a_I$ (whose lowest component is the gaugino), $U_{I\a}{}^\b$
(whose lowest component is related to the gauge field strength)
and $\O_{A\a}{}^\b$ is the Lorentz connection.

The action \action\ has two local Lorentz transformations. One
rotates the spinor indices as $\d\l^\a=\l^\b\S_\b{}^\a$ and the
other rotates the vector indices as $\d\Pi^a=\Pi^b\L_b{}^a$. At the
end of the day, both Lorentz transformations turn out to be
identified, namely $tr(\g^{ab}\S)$ is proportional to $\L^{ab}$. The
action is also invariant under gauge transformations of the gauge
group $SO(32)$ or $E_8\times E_8$.

The quantization of the system given by \action\ is performed by
studying the cohomology of the BRST operator $Q_{BRST}=\oint
j_{BRST} = \oint \l^\a d_\a$. As it was demonstrated in \bh, the
constraints on the backgrounds fields of the ten dimensional
SUGRA/SYM system are implied by the nilpotence of the BRST charge
and the holomorphicity of the BRST current. Let us reobtain this
result in slightly different manner. We first determine how the
world-sheet fields are transformed by the action of $Q_{BRST}$. In
order to do this, we define the canonical conjugate to $Z^M$ as

\eqn\mom{P_M = (2\pi\a') { {\d S} \over {\d(\p_0 Z^M)} } } $$ =
-E_M{}^\a d_\a + \ha ( \Pi_a + \Pb_a ) E_M{}^a - \ha ( \Pi^A -
\Pb^A ) B_{AM} + \Jb^I A_{IM} + \l^\a \o_\b \O_{M\a}{}^\b,$$ where
$\p_0$ is respect to the world-sheet time $\s^0$. In this way we
can relate $P_M$ to the world-sheet field $d_\a$. We also use the
canonical commutation relations

\eqn\comm{[ P_M, Z^N ] = -\d^N_M,\quad [ \l^\a, \o_\b ] =
\d^\a_\b,\quad [\Jb^I, \Jb^J ] = f^{IJ}{}_K \Jb^K,} where $f$'s are
structure constants of the gauge group. Note that these commutations
relations are done at equal world-sheet times and that there is a
delta function $\d(\s^1-\s'^1)$ in the r.h.s. of each.

As it was shown in \bh, the nilpotence of $Q_{BRST}$ can be
computed after writing $d_\a$ in terms of the canonical variables
and using the canonical commutations relations of \comm. The
holomorphicity of $j_{BRST}$ is determined from the equations of
motion derived from the action \action. In these ways, the
background fields satisfy the nilpotence constraints

\eqn\qq{\l^\a \l^\b T_{\a\b}{}^A = \l^\a \l^\b H_{\a\b A} = \l^\a
\l^\b F_{I\a\b} = \l^\a \l^\b \l^\g R_{\a\b\g}{}^\d =0,} and the
holomorphicity constraints

\eqn\dj{T_{\a(ab)} = H_{\a ab} = T_{a\a}{}^\b = \l^\a \l^\b
R_{a\a\b}{}^\g = T_{\a\b a} - H_{\a\b a} = F_{Ia\a} - W^\b_I T_{\a\b
a} = 0,}
$$
F_{I\a\b} - \ha H_{\a\b\g}W^\g_I = \N_\a W^\b_I + W^\g_I
T_{\g\a}{}^\b - U_{I\a}{}^\b = \l^\a\l^\b( \N_\a U_{I\b}{}^\g +
R_{\d\a\b}{}^\g W^\d_I ) =0,$$ where $T, R, H$ and $F$ are the
torsion, the Lorentz curvature, the gauge field strength and the
three-from curvature of the two-form $B$. In \bh\ it was proved
that these constraints put the background fields on-shell, that
they satisfy the N=1 D=10 SUGRA/SYM equations of motion.

\newsec{BRST Transformations of the World-sheet Fields}

We define the BRST transformation of a field $\Psi$ as
$\d_B\Psi=[\oint \e\l^\a d_\a, \Psi]$, where $\e$ is a constant
Grassmann number and the Poisson bracket is calculated from the
canonical commutation relations of \comm. To do this, we need to
express the world-sheet field $d_\a$ in terms of the $P_M$ and the
other world-sheet fields. From the definition \mom\ one obtains

\eqn\ddep{d_\a = -E_\a{}^M P_M - \ha ( \Pi^A - \Pb^A ) B_{A\a} +
\Jb^I A_{I\a} + \l^\b \o_\g \O_{\a\b}{}^\g.}

Now it will be shown that $\d_B=\d_g+\td$, where $\d_g$ refers to
the gauge transformation with parameter $-\e\l^\a A_{I\a}$ and a
Lorentz transformation with parameter $-\e\l^\g \O_{\g\b}{}^\a$.

Consider first the pure spinor $\l^\a$. We obtain

$$
\d_B \l^\a = \oint d\s' \e\l^\b(\s') [ d_\b(\s') , \l^\a(\s) ] =
\oint d\s' \e\l^\b(\s') \l^\g(\s') \O_{\b\g}{}^\d(\s') [
\o_\d(\s') , \l^\a(\s) ]$$ $$ =  \l^\b (-\e \l^\g
\O_{\g\b}{}^\a),$$ which corresponds to a Lorentz rotation of the
pure spinor $\l^\a$ with parameter $-\e\l^\g \O_{\g\b}{}^\a$.

Consider now the conjugate pure spinor $\o_\a$. Its BRST variation
becomes

$$
\d_B \o_\a = \oint d\s' \e [ \l^\b d_\b (\s') , \o_\a(\s) ] =
\oint d\s' \e \l^\b (\s') [ d_\b(\s') , \o_\a(\s) ] + \e [
\l^\b(\s') , \o_\a(\s) ] d_\b(\s')$$ $$ = -(-\e\l^\b
\O_{\b\a}{}^\g) \o_\g + \e d_\a,
$$
where the first term is a Lorentz rotation.

Consider now the BRST variation of the gauge current $\Jb^I$. It
is given by

$$
\d_B \Jb^I = \oint d\s'\e \l^\a (\s') [ d_\a(\s') , \Jb^I(\s) ] =
f^{IJ}{}_K (-\e\l^\a A_{J\a}) \Jb^K,$$ which is a gauge
transformation in the adjoint representation with $-\e\l^\a
A_{J\a}$ as gauge parameter.

Now we consider the BRST transformation of $\Pi^A$. To obtain them
we note that $\d_B Z^M = \e \l^\a E_\a{}^M$, then

$$
\d_B \Pi^A = \d_B ( \p Z^M E_M{}^A ) = \p ( \d_B Z^M E_M{}^A ) -
\d_B Z^M \p Z^N \p_{[N} E_{M]}{}^A,$$ if we use the definition of
the torsion, then we have

$$
\d_B \Pi^A = \p ( \e \l^\a \d_\a^A ) + \e \l^\a \Pi^A \O_{A\a}{}^A
- \e \l^\a \Pi^B T_{B\a}{}^A - \e \l^\a \Pi^B \O_{\a B}{}^A
(-1)^B.
$$
Therefore,

\eqn\pis{\d_B \Pi^a =  \Pi^b( -\e \l^\a \O_{\a b}{}^a )  - \e
\l^\a \Pi^B T_{B\a}{}^a,\quad \d_B \Pi^\a = \Pi^\b ( -\e \l^\g
\O_{\g\b}{}^\a ) + \N ( \e \l^\a ) - \l^\b \Pi^B T_{B\b}{}^\a,}
where the first term in each transformation corresponds to a
Lorentz rotation and $\N ( \e \l^\a ) = \p ( \e \l^\a ) + \e\l^\b
\Pi^A \O_{A\b}{}^\a$. We obtain analogous transformations for
$\Pb^A$.

The BRST transformation of any background superfield is given by
$\d_B \Psi = \e \l^\a \p_\a \Psi$. It can be shown that this
expression can also be written as a gauge transformation for
$\psi$ plus a term which depends on the covariant derivative of
the superfield. For example, for the superfield $W^\a_I$ one
obtains

\eqn\dwi{\d_B W^\a_I = W^\b_I ( -\e\l^\g\O_{\g\b}{}^\a ) -
f^{JK}{}_I ( - \e\l^\b A_{J\a} ) W^\a_K + \e \l^\b \N_\b W^\a_I,}
where the first term is Lorentz rotation of $W^\a_I$ and the
second is a gauge transformation of $W^\a_I$.

\subsec{Nilpotency}

Now it will be shown that $\d_B^2$ acting on the world-sheet
fields leads to the nilpotence constraints of \qq. Consider $Z^M$
first

$$
\d_B^2 Z^M = \d_B (\e_1 \l^\a E_\a{}^M) = \e_1 ( \d_B \l^\a )
E_a{}^M + \e_1 \l^\a \d_B E_\a{}^M  = \e_1\e_2 \l^\a \l^b ( -
\O_{\a\b}{}^\g E_\g{}^M - \p_\b E_\a{}^M ),
$$
by symmetrizing in $(\a\b)$ we form the torsion $T_{\a\b}{}^A
E_A{}^M$. Therefore we obtain the constraint $\l^\a \l^\b
T_{\a\b}{}^A = 0$.

Similarly, we compute $\d_B^2 \l^\a$

$$
\d_B^2 \l^\a = \d_B ( - \e_1 \l^\b \l^\g \O_{\g\b}{}^\a ) = - \e_1
\e_2 \l^\b \l^\g \l^\d ( \p_\d \O_{\g\b}{}^\a- \O_{\b\g}{}^\s
\O_{\d\s}{}^\a - \O_{\g\d}{}^\s \O_{\s\b}{}^\a ),$$ after
symmetrizing in $(\b\g\d)$ we form the curvature components
$R_{\d\g\b}{}^\a$, then we obtain the constraint $\l^\b \l^\g
\l^\d R_{\d\g\b}{}^\a = 0$.

Now we consider the gauge current $\Jb^I$

$$
\d_B^2 \Jb^I = \d_B ( -\e_1 f^{IJ}{}_K \l^\a A_{J\a} \Jb^K ) =
-\e_1 \e_2 \l^\a \l^\b f^{IJ}{}_K \Jb^K ( \p_\b A_{J\a} -
\O_{\a\b}{}^\g A_{J\g} )$$ $$ + \e_1 \e_2 \l^\a \l^\b f^{IJ}{}_K
f^{KL}{}_M A_{J\a} A_{L\b} \Jb^M,$$ if we symmetrize in $(\a\b)$
and use the fact that the structure constants are the group
generators in the adjoint representation, then we can form the
field-strength $F_{I\a\b}$ and we obtain the constraint $\l^\a
\l^\b F_{I\a\b} = 0$. It remains to check the nilpotence
constraint for the superfield $H$. For this we need to transform
$d_\a$ under the pure spinor BRST charge.

\subsec{BRST transformation of the superspace translations
generator}

Now we consider the world-sheet field $d_\a$. Its BRST variation
is given by

$$
\d_B d_\a = \oint \e ( - [ \l^\b(\s') , d_\a(\s) ] d_\b(\s') +
\l^\b(\s') [ d_\b(\s') , d_\a(\s) ] )$$ $$ = - \e \l^\g
\O_{\a\g}{}^\b d_\b  + \oint d\s' \e \l^\b(\s') [ d_\b(\s') ,
d_\a(\s) ],$$ the first term here is not a Lorentz rotation as it
was promised. The Lorentz rotation term will appear after the
computation of the second term. To do this, we remind the relation
between  $d_\a$ and the remaining world-sheet field \ddep. The
more difficult brackets to compute are those coming from the first
terms in \ddep. It is due to the fact that there will appear some
part integrations to get the right result. After doing the other
commutators we obtain

$$
\oint d\s' \e \l^\b(\s') [ d_\b(\s') , d_\a(\s) ] = \e \l^\b [ - (
E_\b{}^M \p_M E_\a{}^N + E_\a{}^M \p_M E_\b{}^N ) P_N + \Jb^I (
\p_{(\a} A_{I\b)} + f^{JK}{}_I A_{J\b} A_{K\a} )$$ $$ + \l^\g
\o_\d ( \p_{(\a} \O_{\b)\g}{}^\d + \O_{\b\rho}{}^\d
\O_{\a\g}{}^\rho - \O_{\b\g}{}^\rho \O_{\a\rho}{}^\d )]$$ $$ +
\oint d\s' \e \l^\b(\s') ( [ E_\b{}^M P_M(\s') , \p_1 Z^N
B_{N\a}(\s) ] + [ E_\a{}^M P_M(\s) , \p_1 Z^N B_{N\b}(\s') ].$$
Let us consider the last integral. After doing the commutators we
get it is equal to

$$
\e\l^\b ( - E_\b{}^M \p_1 Z^N \p_M B_{N\a} (-1)^{MN} - E_\a{}^M
\p_1 Z^N \p_M B_{N\b} (-1)^{MN} )$$ $$ - \oint \e\l^\b(\s') (
E_\b{}^M(\s') B_{M\a}(\s) {\p\over{\p\s}} \d(\s-\s') +
E_\a{}^M(\s) B_{M\b}(\s') {\p\over{\p\s'}}\d(\s-\s') ),$$ after
integration on $\s'$ we obtain that this expression becomes

$$
\e\l^\a ( \p_1 Z^M E_\b{}^M \p_M B_{\a N} + \p_1 Z^M E_\a{}^M \p_M
B_{\b N} )$$ $$ + \e ( - (\p_1 \l^\b) B_{\b\a} - \l^\b ( \p_1
E_\b{}^M ) B_{M\a} + ( \p_1 \l^\b ) B_{\a\b} + \l^\b E_\a{}^M \p_1
B_{M\b} )$$
$$
= \e\l^\b ( (-1)^{M+1} E_\b{}^M E_\a{}^P \p_{[P} E_{M]}{}^A \p_1
Z^N B_{NA} + \p_1 Z^N H_{\b\a N} ),$$ where $H$ stands for the
components of the three-form field strength of the two-form
superfield $B$, that is, $H=dB$. Adding up all the contributions
we obtain

$$
\oint d\s' \e [ \l^\b d_\b(\s') , d_\a(\s) ] = \e \l^\b [
(-1)^{M+1} E_\b{}^M E_\a{}^P \p_{[P} E_{M]}{}^A ( E_A{}^N P_N +
\p_1 Z^N B_{NA} ) + \p_1 Z^N H_{\b\a N}$$ $$ + \Jb^I ( \p_{(\a}
A_{I\b)} + f^{JK}{}_I A_{J\b} A_{K\a} ) + \l^\g \o_\d ( \p_{(\b}
\O_{\a)\g}{}^\d + \O_{\b\rho}{}^\d \O_{\a\g}{}^\rho -
\O_{\b\g}{}^\rho \O_{\a\rho}{}^\d ) ].$$ After using

$$
\p_{(\a} A_{I\b)} + f^{JK}{}_I A_{J\b} A_{K\a} = F_{I\b\a} -
(-1)^{M+1} E_\b{}^M E_\a{}^P \p_{[P} E_{M]}{}^A A_{IA},$$
$$
\p_{(\b} \O_{\a)\g}{}^\d + \O_{\b\rho}{}^\d \O_{\a\g}{}^\rho -
\O_{\b\g}{}^\rho \O_{\a\rho}{}^\d = R_{\b\a\g}{}^\d - (-1)^{M+1}
E_\b{}^M E_\a{}^P \p_{[P} E_{M]}{}^A \O_{A\g}{}^\d,$$ (with $F$ is
the gauge field-strength , $R$ is the Lorentz curvature) and
reminding the definition \mom\ we arrive to the BRST
transformation of the world-sheet field $d_\a$ to be

$$
\d_B d_\a = -\e \l^\g \O_{\a\g}{}^\b d_\b + \e \l^\b (-1)^{M+1}
E_\b{}^M E_\a{}^P \p_{[P} E_{M]}{}^A ( \ha ( \Pi_a + \Pb_a )
\d^a_A - \d^\g_A d_\g )$$ $$ + \e\l^\b ( \p_1 Z^N H_{\b\a N} +
\Jb^I F_{I\b\a} + \l^\g \o_\d R_{\b\a\g}{}^\d ),$$ we recall that
the combination of supervielbein appearing in the second term
above is related to the torsion we finally obtain

\eqn\dd{\d_B d_\a = -( - \e\l^\g \O_{\g\a}{}^\b ) d_\b + \e \l^\g
d_\b T_{\g\a}{}^\b + \e \l^\b \l^\g \o_\d R_{\a\b\g}{}^\d +
\e\l^\b \Pi^a T_{\b\a a},} where we recognize a Lorentz rotation
in the first term. Here we need that $F_{I\a\b} = H_{\a\b\g} =
H_{\a\b a} - T_{\a\b a} = 0$ which are consistent with the
constrains derived in \bh. In this way the nilpotence constraint
for $H$ in \qq\ are satisfied.

In summary, we have proved that the BRST transformations contain a
term which corresponds to a gauge and/or Lorentz transformation
with field dependent parameters.

\newsec{BRST Variation of the Action}

As a check we will vary the action \action\ under the
transformations we derived above to derive the holomorphic
constraints of \dj. Before this, let us compute the transformation
of the gauge connection $A_I = \Pi^A A_{IA}$ and that of the
Lorentz connection $\O_\a{}^\b = \Pi^A \O_{A\a}{}^\b$ and after
that we can deduce analogous transformations for $\Ab_I = \Pb^A
A_{IA}$ and $\Ob_\a{}^\b = \Pb^A \O_{A\a}{}^\b$. These
transformations are similar to those of $\Pi^A$, the difference is
that the result does not contain the torsion but the corresponding
curvature. That is, for $A_I$ it will appear the field strength
$F$ and for $\O_\a{}^\b$ it will appear the curvature $R$. The
result is

\eqn\conn{\d_B A_I = - \N( - \e \l^\a A_{I\a} ) - \e \l^\a \Pi^A
F_{IA\a},\quad \d_B \O_\a{}^\b = - \N ( -\e \l^\g \O_{\g\a}{}^\b )
- \e \l^\g \Pi^A R_{A\g\a}{}^\b,} where we recognize the gauge and
Lorentz rotation parts in each transformation.

Since the action is invariant under gauge and Lorentz rotations,
we do not need to include that gauge and Lorentz parts in the BRST
transformations of the fields appearing in \action. Up to gauge
and Lorentz transformations the different terms in \action\
transform in the following way. The variation of the first term is
proportional to

$$
\int d^2z \e \l^\a \Pi^{(A} \Pb^{a)} T_{A\a a}.$$ The variation of
the second term of the action is proportional to

$$
\int d^2z \e \l^\a \Pi^A \Pb^B H_{BA\a},$$ here we have performed
integrations by parts and the identity $\Nb \Pi^A - \N \Pb^A =
\Pi^B \Pb^C T_{CB}{}^A$. The variation of the third term is
proportional to

$$
\int d^2z - \e \l^\a \Pi^A \Jb^I F_{IA\a}.$$ The variation of the
fourth term of the action is proportional to

$$
\int d^2z \e [ - d_\a \Nb \l^\a - \l^\a d_\b \Pb^\g T_{\a\g}{}^\b
+ \l^\a \l^\b \o_\g \Pb^\d R_{\a\d\b}{}^\g + \l^\a \Pi^a \Pb^\b
T_{\a\b a} + \l^\a d_\b \Pb^A T_{A\a}{}^\b ].$$ The variation of
the fifth term is

$$
\int d^2z \e [ \l^\a d_\b \Jb^I T_{\a\g}{}^\b W^\g_I + \l^\a \l^\b
\o_\g \Jb^I R_{\d\a\b}{}^\g W^\d_I + \l^\a \Pi^a \Jb^I T_{\a\b a}
W^\b_I - \l^\a d_\b \Jb^I \N_\a W^\b_I ].$$ The variation of the
sixth term plus the $S_{\l, \o}$ is

$$
\int d^2z \e [ d_\a \Nb \l^a - \l^\a \l^\b \o_\g \Pb^A
R_{A\a\b}{}^\g ].$$ The variation of the seventh term is

$$
\int d^2z \e [ \l^\a d_\b \Jb^I U_{I\a}{}^\b + \l^\a \l^\b \o_\g
\Jb^I \N_\a U_{I\b}{}^\g ].$$ And the variation of $S_{\Jb}$ is
zero up to gauge transformation.

After summing up all the contributions, we note that the terms
involving $\Nb\l^\a$ and $\Pb^\a$ are zero. Finally the variation
of the action becomes

\eqn\qs{ \d_B S = {1\over{2\pi\a'}} \int d^2z \e [ \half \l^\a
\Pi^a \Pb^b ( T_{\a(ab)} + H_{ba\a} ) + \half \l^\a \Pi^\b \Pb^a (
H_{\b\a a} - T_{\b\a a} ) + \l^\a d_\b \Pb^a T_{a\a}{}^\b} $$
 - \l^\a \l^\b
\o_\g \Pb^a R_{a\a\b}{}^\g + \l^\a \Pi^a \Jb^I ( \half( H_{\a\b a} +
T_{\a\b a} ) W^\b_I - F_{Ia\a} ) + \l^\a \Pi^\b \Jb^I ( \half
H_{\a\b\g} W^\g_I - F_{I\a\b} )
$$
$$ + \l^\a d_\b \Jb^I ( U_{I\a}{}^\b+ T_{\a\g}{}^\b
W^\g_I - \N_\a W^\b_I )
 + \l^\a \l^\b \o_\g \Jb^I ( \N_\a
U_{I\b}{}^\g + R_{\d\a\b}{}^\g W^\d_I) ].$$ Therefore, the
condition $\d_B S = 0$ determines the classical constraints \dj\
on the background superfields.

\vskip 15pt {\bf Acknowledgements:} I would like to thank Nathan
Berkovits for useful comments. This work is supported by
Fundaci\'on Andes, FONDECYT grant 1061050 and Proyecto Interno
27-05/R from UNAB.

\listrefs

\end